\documentclass[reqno,12pt]{article}
\usepackage{amsmath,amsfonts,amssymb,amsthm,amstext,amscd,eucal}
\usepackage[all]{xy}
\usepackage{hyperref}
\usepackage{epsfig}\usepackage{graphicx}

\makeatletter \@addtoreset{equation}{section}

\makeatletter\renewcommand\section{\@startsection {section}{1}{\z@}%
                                   {-3.5ex \@plus -1ex \@minus -.2ex}
                                   {2.3ex \@plus.2ex}%
                                   {\normalfont\large\bfseries}}
\renewcommand\subsection{\@startsection{subsection}{2}{\z@}%
                                     {-3.25ex\@plus -1ex \@minus -.2ex}%
                                     {1.5ex \@plus .2ex}%
                                     {\normalfont\bfseries}}

\newcommand{\eg}{{\em e.g. }}

\newcommand{\be}{\begin{equation}}
\newcommand{\ee}{\end{equation}}
\newcommand{\bea}{\begin{eqnarray}}
\newcommand{\eea}{\end{eqnarray}}

\newcommand{\bse}{\begin{subequations}}
\newcommand{\ese}{\end{subequations}}
\newcommand{\bi}{\begin{itemize}}
\newcommand{\ei}{\end{itemize}}

\newcommand{\dl}{\delta\Lambda(t)}

\begin{document}
\begin{titlepage}
\begin{flushright}\vspace{-3cm}
{\small
IPM/P-2012/008 \\
\today }\end{flushright}\vspace{3mm}

\begin{center}

\centerline{{\Large{\bf Revisiting Cosmic No-Hair Theorem for Inflationary Settings}}}
\vspace{5mm}

{\large{\bf {A.~Maleknejad$^{a,b,}$\footnote{azade@ipm.ir}, M.M. Sheikh-Jabbari$^{a,}$\footnote{jabbari@theory.ipm.ac.ir}}}}

\bigskip\medskip

$^a$ School of Physics, Institute for research in fundamental sciences
(IPM),\\ P.O.Box 19395-5531, Tehran, Iran\\
\vspace{1mm}
$^b$ Department of Physics,
Alzahra University
P. O. Box 19938, Tehran 91167, Iran

\end{center}

\vspace{8mm}
\setcounter{footnote}{0}


\begin{abstract}
\noindent In this work we revisit Wald's cosmic no-hair theorem \cite{Wald} in the context of accelerating Bianchi cosmologies for a generic cosmic fluid with non-vanishing anisotropic stress tensor and when the fluid energy momentum tensor is  of the form of a cosmological constant term plus a piece which does \emph{not} respect strong or dominant energy conditions. Such a fluid is the one appearing in inflationary models. We show that for such a system anisotropy may grow, in contrast to the cosmic no-hair conjecture. In particular, for a generic inflationary model we show that there is an upper bound on the growth of anisotropy. For slow-roll inflationary models our analysis can be refined further and the upper bound is found to be of the order of slow-roll parameters. We examine our general discussions and our extension of Wald's theorem for three classes of slow-roll inflationary models, generic multi-scalar field driven models, anisotropic models involving U(1) gauge fields and the gauge-flation scenario.

\end{abstract}


\vspace{0.5in}

\end{titlepage}


\renewcommand{\baselinestretch}{1.05}  

\tableofcontents

\section{Introduction}

The Universe at cosmological scales and as we see it today looks homogenous and isotropic. To understand this almost perfectly homogenous and isotropic Universe, given that the cosmic evolution may start from a generic initial condition over which we have no control, it is natural to seek a ``dynamical'' explanation: isotropic and homogenous Universe is an attractor of the cosmic evolution. According to the standard model of cosmology  cosmic evolution is governed by the Einstein gravity coupled to the cosmic fluid, this latter idea if true, should be an outcome of  Einstein equations. The first such attempt was made in \cite{Gibbons-Hawking,Hawking-Moss} arguing that the late-time behavior of any  accelerating Universe  is an isotropic Universe. This statement was dubbed as ``cosmic no-hair conjecture''. The first attempt to prove this conjecture was presented in Wald's seminal paper \cite{Wald}. Wald's cosmic no-hair theorem  states that Bianchi-type models (except Bianchi IX) with the total  energy momentum tensor of the form $T_{\mu\nu}=-\Lambda_0 g_{\mu\nu}+ \tilde T_{\mu\nu}$, and with a constant positive $\Lambda_0$ and a $\tilde T_{\mu\nu}$ satisfying Strong and Dominant Energy Conditions, respectively SEC and DEC, will approach de Sitter space exponentially fast, within a few Hubble times $H^{-1}=\sqrt{3/\Lambda_0}$.  For a more thorough historical review and other related works on cosmic no-hair conjecture/theorem see \cite{Math-cosmology} and references therein.

Cosmological observations, specially the CMB data \cite{CMB-data}, indicate that the early Universe have in fact gone through such an accelerated expansion inflationary period. As a result, if cosmic no-hair conjecture/theorem holds,
all traces of  initial anisotropy and inhomogeneity should be washed away \cite{Inflation-texts}. On the other hand, inflation has ended and we are not in a de Sitter Universe  with Hubble parameter equal the one at the end of inflation, as Wald's theorem suggests. This is simply because inflationary setups are constructed to be so and do not strictly respect SEC or DEC assumptions of Wald's theorem. (For a more rigorous discussion of the latter fact see appendix B.) Cosmic no-hair theorem should hence be revisited for inflationary models. Such studies has of course been carried out extensively, e.g. see \cite{Sahni} and references in \cite{Math-cosmology} for an incomplete list. It may still seem plausible to expect that the result of Wald's theorem, namely an isotropic Universe, to hold for inflationary models. This is due to the fact that inflation is expected to have lasted around 60 e-folds, i.e. 60 Hubble times, while the time scale for anisotropy damping, if Wald's theorem holds, is a few Hubble times.\footnote{As we will discuss in sections 3 and 4, this argument may be used for cases where the energy momentum of the system driving inflation does not contribute to anisotropic stress, like scalar-driven model.}

The above expectation, and applicability  of cosmic no-hair conjecture for inflationary models, may be challenged in two different ways: by changing the setup upon which Wald's theorem is based. That is, modifying gravity theory used for inflationary model building; or alternatively, to search for particular inflationary models which drastically violate SEC or DEC assumptions. Both of these ways have been studied, e.g. see \cite{Barrow} for the former and \cite{Vector-anisotropy,Kanno} for the latter. All of our analysis in this work will be within Einstein gravity and hence applicable to the latter class of models.

As pointed out and will be discussed briefly in section 3 and in more details in appendix B, during inflation assumptions of Wald's theorem are violated. However, this violation should be such that the model allows for a long enough inflationary period and has viable observational signatures. It has been shown that it is indeed possible to construct such stable anisotopic inflationary models, e.g.  see \cite{Kanno,Himmetoglu,Hervik}. There is hence a need to revisit the fate of Wald's theorem in more general inflationary setups and in particular slow-roll models. This is the question we tackle in this work.

In this work, in section 2, we study evolution of Bianchi cosmological models within Einstein gravity sourced by
cosmic fluid with anisotropic stress. As in Wald's work \cite{Wald} we decompose the energy momentum tensor as $-\Lambda(t) g_{\mu\nu}+\mathcal{T}_{\mu\nu}$, with $\Lambda>0$ and $\mathcal{T}_{\mu\nu}$ satisfying SEC and WEC, but now allow for $\Lambda$ to have time dependence.
As proved in appendix A, generic inflationary models conform to this decomposition. In section 3, we push our analysis further by concentrating on inflationary frameworks, and in particular slow-roll, quasi-de Sitter expansion. This allows us to refine our results, obtain an upper bound and study the time evolution of the anisotropy. In section 4, we apply our general arguments of sections 2 and 3 to two classes of models: scalar-driven inflationary models and inflationary models with background gauge fields. In the last section we summarize our results into the ``inflationary extended cosmic no-hair theorem'' and briefly discuss the observational consequences of our results.

\section{Extended cosmic no-hair, the general setup}
The Bianchi family are cosmological models with spatially homogeneous surfaces which are invariant under the action of a  three dimensional Lie group, called the symmetry group. A Bianchi model can always be written as (e.g. see \cite{MacCallum-Stephani})
\be\label{metr}
ds^2=-dt^2+h_{ij}e^i\otimes e^j,
\ee
where $i,j=1,2,3$, label the coordinates in homogeneous space-like hypersurfaces $\Sigma_t$ and $h_{ij}=h_{ij}(t)$ is the spatial metric which can be decomposed as
\be
h_{ij}=e^{2\alpha}e^{2\beta_{ij}}.
\ee
Here $e^{\alpha}$ is the isotropic scale factor and $\beta_{ij}$ is a traceless matrix which describes the anisotropies.
Furthermore, $e^i$ are characterizing one-forms with the following property
\be\label{cartan}
\textmd{d}e^i=-\frac12c^i_{~jk}e^j\wedge e^k,
\ee
where $c^i_{~jk}$ are the structure constants of the corresponding Bianchi type.

Let $n^\mu$ be the unit tangent vector field of the congruence of timelike geodesics orthogonal to the homogeneous space-like hypersurfaces $\Sigma_t$. Then, we obtain the following covariant form for the spatial metric $h_{\mu\nu}$ \eqref{metr}
\be
h_{\mu\nu}=g_{\mu\nu}+n_\mu n_\nu,
\ee
where $g_{\mu\nu}$ is the metric of the space-time.
The extrinsic curvature of $\Sigma_t$ is defined as
\be
K_{\mu\nu}\equiv \frac12\mathfrak{L}_n h_{\mu\nu}=\frac12\dot{h}_{\mu\nu}\,,
\ee
where the dot represents  derivative with respect to the proper time $t$.
One can decompose $K_{\mu\nu}$ into  trace, and  trace-free parts
\be\label{decomk}
K_{\mu\nu}=\frac13 Kh_{\mu\nu}+\sigma_{\mu\nu}\,,
\ee
where
\be\label{hubble}
K(t)=K_{\mu\nu}h^{\mu\nu}=3H(t)\,.
\ee
Here $H(t)=\dot\alpha$ is the Hubble parameter corresponding to the homogenous scale factor $e^{\alpha}$. The shear of the time-like geodesic congruences $\sigma_{\mu\nu}$ is related to the time derivative of $\beta_{ij}$ and is a symmetric, traceless and purely spatial tensor:
\be\label{prosigma}
h_{\mu\nu}\sigma^{\mu\nu}=n_\mu\sigma^{\mu\nu}=0\,.
\ee

We now analyze Einstein's equation for Bianchi models
\be\label{einst}
G_{\mu\nu}=T_{\mu\nu}=-\Lambda(t) g_{\mu\nu}+\mathcal{T}_{\mu\nu},
\ee
where $\Lambda(t)$ is a positive but time-dependent cosmological term and in our conventions we set $8\pi G=1$. We assume that $\mathcal{T}_{\mu\nu}$ satisfies the strong and weak energy conditions. It is shown in the appendix A that it is always possible to decompose energy momentum tensor of any inflationary system in this way.
The strong energy condition (SEC) states
\be\label{secT}
(\mathcal{T}_{\mu\nu}-\frac12\mathcal{T}g_{\mu\nu})t^\mu t^\nu\geq0,\qquad\textmd{ for all time-like $t^\mu$}.
\ee
The dominant energy condition (DEC) stipulates that \be\label{WEC}
\mathcal{T}_{\mu\nu}t^\mu t'^\nu\geq0\ \quad\textmd{for all future-directed causal vectors $t^\mu, t'^\nu$}.
\ee
We note that the above for $t'^\mu=t^\mu$ leads to weak energy condition (WEC) $\mathcal{T}_{\mu\nu}t^\mu t^\nu\geq0$.

One can then decompose Einstein equations \eqref{einst} into  four constraint equations
\bea
\label{initialvalue}
\mathcal{T}_{\mu\nu}n^\mu n^\nu&=&\frac12{}^{^{(3)\!\!}}R-\frac12\sigma_{\mu\nu}\sigma^{\mu\nu}+\frac13K^2-\Lambda(t)\,,\\
\label{constraint}
\mathcal{T}_{\sigma\lambda}h^\sigma_{~i}n^\lambda&=&K^\sigma_{~\lambda}c^\lambda_{~\sigma i}+K^\sigma_{~i}c^\lambda_{~\sigma\lambda}\,,
\eea
and six dynamical equations
\bea
\label{dyneinst}
(\mathcal{T}_{\sigma\lambda}-\frac12h_{\sigma\lambda}\mathcal{T})h^\sigma_{~i}h^\lambda_{~j}=\mathfrak{L}_n\sigma_{ij}+\frac13(\dot Kh_{ij}+K^2h_{ij}+K\sigma_{ij})-2\sigma_{i\lambda}\sigma^\lambda_{~j}-\Lambda(t)h_{ij}+{}^{^{(3)\!\!}}R_{ij}\,.~~~~~~~
\eea
Here ${}^{^{(3)\!\!}}R_{ij}$ is the spatial Ricci tensor and can be written in terms of the structure-constant tensor $c^i_{jk}$ \eqref{cartan} as
\bea
{}^{^{(3)\!\!}}R_{ij}=\frac14c_{ikl}c_j^{~kl}-c^k_{~kl}c_{(ij)}^{~~~l}-c_{kli}c^{(kl)}_{~~j}\,,
\eea
where indices raised and lowered with metric $h_{ij}$, and ${}^{^{(3)\!\!}}R={}^{^{(3)\!\!}}R_{ij}h^{ij}$ is the spatial curvature of $\Sigma_t$.
Note that all but one of Bianchi models have  negative spatial curvature
\be\label{3-curvature}
{}^{^{(3)\!\!}}R\leq0\,;
\ee
Bianchi type IX has a positive curvature ${}^{^{(3)\!\!}}R\geq0$ \cite{Wald,{MacCallum-Stephani}}.

Combining trace of \eqref{dyneinst} with \eqref{initialvalue}, we obtain Raychaudhuri equation
\be \label{ray}
(\mathcal{T}_{\mu\nu}-\frac12g_{\mu\nu}\mathcal{T})n^\mu n^\nu=-\dot
K+\Lambda(t)-\frac13K^2-\sigma_{\mu\nu}\sigma^{\mu\nu},
\ee
where $\dot K\equiv\mathfrak{L}_n K$. Contracting \eqref{dyneinst} with $h^{ij}$ and removing the trace part, we have the following equation for the shear tensor
 \be\label{anisotropy}
 \dot{\sigma}^i_{~j}+K\sigma^i_{~j}+{}^{^{(3)\!\!}}S^{i}_{~j}= \mathcal{T}_{kl}h^{ki}h^l_{~j}-\frac13\mathcal{T}_{kl}h^{kl}h^i_{~j}\,.
 \ee
 where ${}^{^{(3)\!\!}}S^{i}_{~j}$ is the anisotropic part of the spatial 3-curvature
 \be
  {}^{^{(3)\!\!}}S^{i}_{~j}={}^{^{(3)\!\!}}R^{i}_{~j}-\frac13\ {}^{^{(3)\!\!}}Rh^i_{~j}\,.
 \ee

Integrating equation \eqref{anisotropy}, we obtain the following integral equation for $\sigma^i_{~j}(t)$
\be\label{anisotInt}
\sigma^i_{~j}(t)=e^{-3\alpha(t)}\int_{t_0} ^t \frac{d(e^{3\alpha(t')})}{K(t')}\big(\mathcal{T}_{kl}h^{ki}h^l_{~j}
-\frac13\mathcal{T}_{kl}h^{kl}h^i_{~j}-{}^{^{(3)\!\!}}S^{i}_{~j}\big)+C_{~j}^ie^{-3\alpha(t)}\,,
\ee
with integration constants $C^i_{~j}$. In order to determine $\sigma^i_{~j}(t)$ we need more information about the energy momentum tensor $\mathcal{T}_{\mu\nu}$.

\section{Extended cosmic no-hair theorem for inflationary backgrounds}

Equations of previous section was written for quite general $\Lambda(t)$ and $\mathcal{T}_{\mu\nu}$, and although we restricted our discussions to $\Lambda(t)\geq 0$ and to $\mathcal{T}_{\mu\nu}$ satisfying SEC and DEC, these conditions were not used. In this section, we  focus on our main interest, inflationary systems and crucially use these conditions and  investigate the dynamics of shear tensor and the evolution of anisotropies during inflation.\footnote{
In the cosmology literature, it is more common to work with the Hubble parameter $H(t)$ instead of $K(t)$ (which is more common in GR literature). For the rest of this work we will use $H(t)$ which from \eqref{hubble} is equal to $\frac13K$.}

Inflation is defined as an epoch in the history of universe in which the scale factor has an accelerating expansion
\be\label{inflation}
\frac{\big(e^\alpha\ddot{\big)}}{e^\alpha}=\dot H+H^2 = H^2(1-\epsilon)\geq0,
\ee
where $\epsilon(t)\equiv -\frac{\dot H}{H^2}$ is a positive and growing quantity and inflation eventually ends when $\epsilon\geq1$.
Although condition \eqref{inflation} concerns the dynamics of isotropic part of the geometry, we argue here that
inflation strongly restricts dynamics of anisotropic part of metric and puts an upper bound on the growth of anisotropies.
Note that in spite of some similarities between our set up and \cite{Wald},  the behavior of these two systems are totally different. The increasing $\epsilon(t)$ which is a generic property of inflationary models, is in contrast with the dynamics described by the Wald's theorem \cite{Wald}, in which $\epsilon(t)$ is a positive quantify that asymptotically approaches zero. More technical details about the differences between these two set ups might be found in Appendix B.

Without loss of generality $\mathcal{T}_{\mu\nu}$ may be decomposed as
\be\label{tildeT}
\mathcal{T}_{\mu\nu}(t)=\tilde\rho(t) n_\mu n_\nu+\tilde P(t) h_{\mu\nu}+\pi_{\mu\nu}(t),
\ee
where $\pi_{\mu\nu}$ is the anisotropic stress tensor which is symmetric, traceless, and purely spatial:
$$\pi_{\mu\nu}h^{\mu\nu}=0\,\,\,\,\&\,\,\,\, \pi_{\mu\nu}n^\mu=0\,.$$
We also define $\Lambda_0\equiv 3H^2_0$, where subscript $0$  denotes the initial value (at the beginning of inflation) and
\be\label{deLambda}
\Lambda(t)=\Lambda_0+\dl\,.
\ee
From the combination of \eqref{initialvalue} and \eqref{ray}, after inserting \eqref{tildeT} and \eqref{deLambda}, we obtain
\bea\label{tildeLam}
\frac{\tilde\rho(t)-\tilde P(t)}{2}+\dl&=&\frac13 {}^{^{(3)\!\!}}R+\dot H-\Lambda_0+3 H^2\,,\\
\label{tilderho}
\frac{\tilde\rho(t)+\tilde P(t)}{2}&=&\frac16 {}^{^{(3)\!\!}}R-\dot H-\frac12\sigma_{\mu\nu}\sigma^{\mu\nu}\,.
\eea
Since $H$ is a decreasing quantity during inflation, from \eqref{tildeLam} we learn that, $\frac{\tilde\rho(t)-\tilde P(t)}{2}+\dl$ is a negative quantity in inflationary models of all Bianchi types, expect type IX. Nonetheless, the spatial curvature ${}^{^{(3)\!\!}}R$ has a time dependence proportional to $e^{-2\alpha}$ and this term is damped quickly in inflationary systems. That is,
${}^{^{(3)\!\!}}R$ is damped after a few e-folds and becomes negligible. Therefore, for all Bianchi models, including  Bianchi type IX, even if  $\frac{\tilde\rho(t)-\tilde P(t)}{2}+\dl$ is not negative to start with, it  becomes negative very quickly.
After a few e-folds ${}^{^{(3)\!\!}}R$ is negligible and we approximately have
\bea\label{1}
-\frac{\dl}{3H^2(t)}-\frac{\tilde \rho(t)-\tilde P(t)}{6H^2(t)}&\simeq&-\frac{\dot H(t)}{3H^2(t)}+\frac{\Lambda_0}{3H^2(t)}-1=\frac{\epsilon(t)}{3}+\frac{H_0^2}{H^2(t)}-1,\\
\label{2}
\frac{\tilde\rho(t)+\tilde P(t)}{2H^2(t)}+\frac{\sigma_{\mu\nu}\sigma^{\mu\nu}}{2H^2(t)}&\simeq&-\frac{\dot H(t)}{H^2(t)}=\epsilon(t).
\eea

On the other hand, $\mathcal{T}_{\mu\nu}$ satisfies SEC which choosing $t_\mu=n_\mu+s_\mu$, where $s_\mu$ is a normalized arbitrary space-like 4-vector normal to $n_\mu$, implies
\be\label{DEC-SEC}
\begin{split}
\tilde\rho(t)+\tilde P(t)+\pi_{ij}(t)s^is^j\geq0,\quad i,j=1,2,3\,,
\end{split}
\ee
where $s^i=s_\mu h^{i\mu}$. Recalling the fact that $\pi_{ij}$ is traceless and the above inequality should be hold for all $s^i$, one obtains\footnote{Although we have not used here, we remind that WEC on $\mathcal T_{\mu\nu}$  imply $\tilde \rho\geq 0$.}
\be\label{upPi}
\frac{|\pi^i_{~j}(t)|}{H^2(t)}\leq -4\frac{\dot H(t)}{H^2(t)}=4\epsilon(t)\,\quad \forall i,j.
\ee
{}From the combination of \eqref{1} and \eqref{2} and noting that \eqref{DEC-SEC} implies $\tilde\rho(t)+\tilde P(t)\geq 0$ we obtain
\be\label{sigma2-bound}
\begin{split}
\sigma_{\mu\nu}\sigma^{\mu\nu}&\leq 2H^2(t)\epsilon(t)\,,\\
\delta\Lambda(t) &\leq -(\tilde\rho (t)+\frac{\sigma_{\mu\nu}\sigma^{\mu\nu}}{2})\leq 0\,.
\end{split}
\ee
The above already implies a weak upper bound on anisotropy
\be\label{sigma-weak-bound}
|\sigma^i_{~j}|\leq H(t)\sqrt{2\epsilon}\,.
\ee
As we will see in the next section assuming slow-roll leads to a stronger bound.

During inflation (on the average) $\epsilon$ is an increasing quantity and this result among other things shows the possibility of growth for the anisotropic part of stress tensor. This is in contrast with behavior of systems described by the Wald's theorem, in which all elements of $\mathcal{T}_{\mu\nu}$, \eg $\pi^i_{~j}(t)$'s, are damped exponentially with a time scale $H_0^{-1}$ \cite{Wald}.

Inserting \eqref{tildeT} into \eqref{anisotInt}, we have the following form for the Hubble-normalized shear tensor, $\frac{\sigma^i_{~j}(t)}{H(t)}$
\be\label{diag}
\frac{\sigma^i_{~j}(t)}{H(t)}=\frac{e^{-3\alpha(t)}}{3H(t)}\int^t_{t_0}{de^{3\alpha(t')}\frac{\pi^i_{~j}(t')-{}^{^{(3)\!\!}}S^i_{~j}(t')}{H(t')}}+C^i_{~j}e^{-3\alpha(t)},
\ee
where $C^i_{~j}$ is a constant at most of the order $\sqrt{\epsilon}$ \eqref{sigma-weak-bound} and  ${}^{^{(3)\!\!}}S^i_{~j}(t)\propto\mathcal{O}(-\dot H)e^{-2\alpha(t)}$. Therefore, the second and the third terms in \eqref{diag} have a damping behavior and soon become negligible. Then, after a few e-folds, we have the following relation for the Hubble-normalized elements of shear tensor
\be\label{diagFinal}
\frac{\sigma^i_{~j}(t)}{H(t)}\simeq\frac{1}{3H(t)e^{3\alpha(t)}}\int^t_{\tilde t_0}{\frac{d(H(t')e^{3\alpha(t')})}{(1+\frac{\dot H(t')}{3H^2(t')})}\ \frac{\pi^i_{~j}(t')}{H^2(t')}}\,,
\ee
where $\tilde t_0$ is a few e-folds after the beginning of inflation. As we see $\frac{\pi^i_{~j}(t)}{H^2(t)}$ acts as a source term and its five degrees of freedom may be determined  for specific  models of inflation. Some examples will be reviewed in section 4.

For models with perfect fluid  $T_{\mu\nu}$, in which $\pi^i_{~j}(t)$ is identically zero, shear tensor $\sigma^i_{~j}(t)$ is  damped exponentially fast. Thus, in this kind of systems inflation washes away any initial anisotropies and isotropizes the system, in accord with the cosmic no-hair conjuncture.
This is \emph{not} necessarily the case in a general inflationary model with non-zero anisotropic stress tensor  $\pi^i_{~j}(t)$.

To summarize,  we have so far studied the dynamics of anisotropies during inflation in a model independent way.
We showed that in presence of anisotropic stress tensor $\pi^i_{~j}$  elements of shear tensor $\sigma^i_{~j}(t)$ \emph{are allowed} to grow by inflationary dynamics, which is against the cosmic no-hair conjuncture. Although $\sigma^i_{~j}(t)$ can increase, as one expects from \eqref{upPi}, inflation enforces an upper bound value on their enlargements.

\subsection{quasi de Sitter expansion}
Thus far we considered a general model of inflation. In the following we focus on the slow-roll models. We  investigate the dynamics of system  assuming that ${\pi^i_{~j}(t)}/{H^2(t)}$ always saturates its maximum value,
and determine the possible upper bound value of anisotropies during slow-roll inflation.

We start with defining the so called slow-roll parameters \footnote{Note that our definition of slow-roll parameters $\epsilon$ and $\eta$ and those usually used for the single scalar FLRW inflationary theory with $\mathfrak{L}=\frac12\dot\varphi^2 - V(\varphi)$,  $\epsilon_\phi=\frac{1}{2}\frac{V'^2}{V^2}$ and $\eta_\phi=\frac{V''}{V}$, are related as $\epsilon_\phi=\epsilon$ and $\eta_\phi=\eta+\epsilon$.}
\be\label{slowrollparameters}
\epsilon=-\frac{\dot H}{H^2}\,,\qquad
\eta=-\frac{\ddot H}{2\dot H H}\,, 
\ee
which during the slow-roll are both very small and almost constant.

From \eqref{tilderho} and \eqref{upPi}, we find that $\mathcal T_{\mu\nu}$, $\sigma_{\mu\nu}\sigma^{\mu\nu}$ and ${}^{^{(3)\!\!}}R$ are at most of the order $\epsilon H^2$.
 Hence recalling \eqref{initialdecom1} and using \eqref{initialvalue}, we obtain
\bea
H(t)&\simeq&H_0\big(1-\epsilon_0H_0t\big),
\eea
where $\simeq$ means to the first order in slow-roll parameters, subscript 0 denotes the initial value and $H_0=\sqrt{\Lambda_0/3}$.  Inserting the above relation in \eqref{1} and \eqref{2}, we obtain \footnote{Note that we are considering simple slow-roll models where $\epsilon(t)$ is an always increasing function during inflation, $\dot\epsilon>0$. From the definition of $\epsilon$ \eqref{slowrollparameters} we learn that $\frac{\dot\epsilon}{H\epsilon}=2(\epsilon-\eta)$.}
\bea\label{sloweq}
\frac{\dl}{H^2}+\frac{\tilde\rho(t)-\tilde P(t)}{2H^2(t)}&\simeq&-\epsilon_0(1+6 H_0 t),\\
\label{freeeq}
\frac{\tilde\rho(t)+\tilde P(t)}{2H^2(t)}+\frac{\sigma_{\mu\nu}\sigma^{\mu\nu}}{2H^2(t)}&\simeq&\epsilon_0\big(1+2(\epsilon_0-\eta_0)H_0t\big).
\eea
As we see in \eqref{freeeq}, $\sigma^{\mu\nu}\sigma_{\mu\nu}$ is not directly related to the slow-roll dynamics, but the combination of $\frac{\tilde\rho(t)+\tilde P(t)}{2H^2(t)}+\frac{\sigma_{\mu\nu}\sigma^{\mu\nu}}{2H^2(t)}$ should be slow varying during inflation. As a result,  $\sigma_{\mu\nu}$ is not fully determined by the slow-roll dynamics, while \eqref{sloweq} implies that the isotropic part of the system is governed by the slow-roll. Thus, assuming slow-roll inflation, anisotropies can still evolve quickly in time.

Applying slow-roll approximation in \eqref{diagFinal}, we obtain the following form for the late time behavior of the Hubble-normalized shear tensor $\frac{\sigma^i_{~j}(t)}{H(t)}$
 \be\label{sigmasl}
\frac{\sigma^i_{~j}(t)}{H(t)}\simeq H_0(1+\epsilon_0H_0t)e^{-3H_0t}\int^t_{t_0}{dt'e^{3H_0t'}
(1-2\epsilon_0H_0t')\frac{\pi^i_{~j}(t')}{H^2(t')}}.
\ee
In order to determine $\frac{\sigma^i_{~j}(t)}{H(t)}$, we need $\pi^i_{~j}(t)$'s five degrees of freedom which should be provided by  independent equations. However, regardless of the particular inflationary model which we consider, recalling \eqref{upPi}, slow-roll enforces an upper bound on the enlargement of anisotropies. Rewriting \eqref{upPi} in terms of slow-roll parameters, we obtain
\be\label{SLupPi}
 \frac{|\pi^i_{~j}(t)|}{H^2(t)} \leq 4\epsilon(t)\simeq4\epsilon_0(1+2(\epsilon_0-\eta_0)H_0t))\,.
 \ee
As mentioned before, from the combination of \eqref{2} and \eqref{diagFinal}, we realize that during the slow-roll inflation, ${\pi^i_{~j}(t)}/{H^2(t)}$ can saturate its upper bound and grow in time.
Then, assuming that ${\pi^i_{~j}}(t)/{H^2(t)}$ always saturates its upper bound during the slow-roll, we can find an upper bound value on the enlargement of ${\sigma^i_{~j}}(t)/{H(t)}$ enforced by the slow-roll dynamics.

Inserting \eqref{SLupPi}
into \eqref{sigmasl}, we obtain the upper bound value on the enlargement of anisotropies at the end of slow-roll inflation
\be\label{upperSigma}
\frac{|\sigma^i_{~j}|}{H}|_{t_{sl}}\leq\frac83(\epsilon_0-\eta_0).
\ee
where $t_{sl}$ is the end of slow-roll inflation and approximately is $H_0\epsilon_0t_{sl}\simeq1$.
Thus, a general model of slow-roll inflation by the end of slow-roll inflation, leaves the cosmos almost isotropic, with a shear of the order $\epsilon_0$. Note that considering slow-roll has improved the bound \eqref{sigma-weak-bound} to \eqref{upperSigma}.

\section{Anisotropy dynamics in two classes of inflationary models}

The discussions of previous sections was made for generic inflationary models. In this section, we consider some examples and discuss the possibility of saturation of the anisotropy upper bound found by specific classes of slow-roll models. In particular we study scalar-driven models and models involving vector gauge fields in the inflationary background.

\subsection{Scalar driven inflationary models}

In this part with in the context of Bianchi cosmology, we investigate the late time behavior of anisotropy in three scalar models of inflation: ordinary multi-scalar field model, K-inflation and DBI inflation.

\begin{itemize}
\item
\textbf{Ordinary multi-scalar field models} with the action \cite{Inflation-texts}
 {\be
 S=\int d^4x\sqrt{-g}\big(\frac{R}{2}-\frac12\partial_\mu\varphi_I\partial^\mu\varphi_I-V(\phi_I)\big),
 \ee
 where $I$ runs from 1 to $N$ and $\varphi_I$'s are $N$ scalar fields,}
\item \textbf{K-inflation} with the action \cite{Inflation-texts}
{\be
S=\int d^4x\sqrt{-g}\big(\frac{R}{2}+P(\varphi,\nabla\varphi)\big),
\ee
where $\varphi$ is a scalar field and $X:=\frac12(\nabla\varphi)^2$,
}
\item \textbf{DBI inflation} with the action \cite{S-T}
{\be
S=\int d^4x\sqrt{-g}\bigg[\frac{R}{2}-\frac1f\left(\sqrt{\mathcal{D}}-1)-V(\varphi^I)\right)\bigg],
\ee
where $\mathcal{D}:=\det(\delta^\mu_\nu+fG_{IJ}\partial^\mu\varphi^I\partial_\nu\varphi^J)$, $I$ runs from 1 to N, $\varphi_I$'s are scalar fields and $G_{IJ}$ is an internal metric which determine the interaction between $\varphi_I$'s.
}
\end{itemize}
The characteristic of  all the above scalar driven models is that their energy momentum tensor $T_{\mu\nu}$ is of the form of a \emph{perfect fluid} and hence the shear stress $\pi(t)^i_{~j}$ is identically zero in these systems. Equation \eqref{diag} then implies that in all of the above multi-scalar driven inflationary models the anisotropy $\sigma_{\mu\nu}$ damps out exponentially fast in a couple of Hubble times, leaving an isotropic background.
These models respect the inflationary no-hair conjuncture.

Note that due to the correspondence between $f(R)$ gravity models and general relativity with a scalar field matter, we expect these systems to show similar behavior in anisotropy damping (regardless of the details of $f(R)$) and hence to respect the inflationary no-hair conjecture.

\subsection{Models of Inflation involving vector gauge fields}
In this part we consider two classes of models with a vector gauge field turned on in the background level. As far as anisotropies are concerned, these models are more interesting than scalar-driven cases. In both of the models, as we will see, the energy momentum tensor has an anisotropic stress $\pi^i_{~j}$ which can source anisotropies. Despite  this fact, the dynamics of the two models is such that anisotropy does not grow in one class of models.

To illustrate a detailed analysis for these models we restrict ourselves to Bianchi type I model, while the results seem to be generic to all Bianchi models. Bianchi type-I axially symmetric metric can be described by the line element
\be\label{axymetric}
ds^2=-dt^2+a(t)^2\big(e^{-4\sigma(t)}dx^2+e^{2\sigma(t)}(dy^2+dz^2)\big),
\ee
where $a(t)$ is the isotropic scale factor and $e^{\sigma(t)}$ represents the anisotropy.
Due to the\emph{ vectorial nature} in these models, energy-momentum tensor has a \emph{non-zero} anisotropic stress $\pi(t)$
\be
T^{\nu}_{~\mu}=diag(-\rho(t), P(t)-\pi(t), P(t)+\frac12\pi(t), P(t)+\frac12\pi(t)),
\ee
which its dynamics (whether it increases or decreases during inflation) determines the fate of anisotropies during inflation.
Note that from \eqref{axymetric}, the shear tensor $\sigma_{\mu\nu}$ and $\dot\sigma$ are related as
\be
\sigma_{\mu\nu}\sigma^{\mu\nu}=6\dot\sigma^2,
\ee
hence here instead of working with $\sigma_{\mu\nu}$, we use $\dot\sigma$.

\subsubsection{Inflationary universe with anisotropic hair}
In \cite{Kanno, Sodas}, M. Watanabe, S. Kanno and J. Soda introduced and studied an inflationary model with anisotropic hair, as a counterexample to the cosmic no-hair conjecture. Motivated from supergravity, their model includes scaler field(s)
as inflaton field(s) coupled to a massless $U(1)$ gauge field.
 The action of the model is given as
\be\label{Jiro'smodel}%
S=\int d^4x\sqrt{-{g}}\left[\frac{R}{2}-\frac12\partial_\mu\varphi\partial^\mu\varphi-V(\varphi)-\frac{1}{4}f^2(\varphi)F_{~\mu\nu}F^{~\mu\nu}\right]\,,
\ee
where $f(\phi)$ is the coupling function of the inflaton field to the vector one and the
field strength of the vector field is given as $F_{\mu\nu}=\partial_\mu A_\nu-\partial_\nu A_\mu.$ Below we review the analysis of \cite{Kanno,Sodas}.

Choosing the temporal gauge, $A_0=0$, homogeneous fields are taken to be of the form $A_\mu=(0, A_x(t), 0, 0)$
and $\varphi=\varphi(t)$. With this ansatz, one can solve  equation of motion for the
vector field as
\be\label{Ax}
\dot A_x=f^{-2}(\varphi)e^{-4\sigma}\frac{P_A}{a(t)}.
\ee
where $P_A$ is an integration constant. Substituting the ansatz and \eqref{Ax} into the action, we obtain the energy density $\rho(t)$, pressure density $P(t)$ and anisotropic stress $\pi(t)$
\bea\label{jiroT}
\rho(t)&=&\frac12\dot\varphi^2+V(\varphi)+\frac12f^{-2}(\varphi)\frac{P^2_{~A}}{a(t)^4}e^{-4\sigma},\nonumber\\
P(t)&=&\frac12\dot\varphi^2-V(\varphi)+\frac16f^{-2}(\varphi)\frac{P^2_{~A}}{a(t)^4}e^{-4\sigma},\nonumber\\
\pi(t)&=&\frac16f^{-2}(\varphi)\frac{P^2_{~A}}{a(t)^4}e^{-4\sigma}.\nonumber
\eea
Note that for this specific model the anisotropic stress $\pi(t)$ is equal to $\frac13\rho_A$, where $\rho_A$ is the contribution of the vector field in the energy density.

Let us consider the chaotic inflation with the following potential
\be
V(\varphi)=\frac12m^2\varphi^2,
\ee
and choose the coupling function to be $f(\varphi)=e^{c\varphi^2/2}$. After inserting $V(\varphi)$ and $f(\varphi)$ into \eqref{jiroT}, one can solve Einstein equations. Being interested in cases in which the energy density of the (gauge) vector field grows during inflation, $c>1$, we realize that the Hubble-normalized anisotropic stress ($\frac{\pi(t)}{H^2}=\frac{\rho_A}{\rho}$) increases in time. As a result, \eqref{diag} enforces the Hubble-normalized shear $\frac{\dot\sigma}{H}$ to grow during inflation.

\begin{figure}[t]
\includegraphics[angle=0,width=75mm, height=65mm]{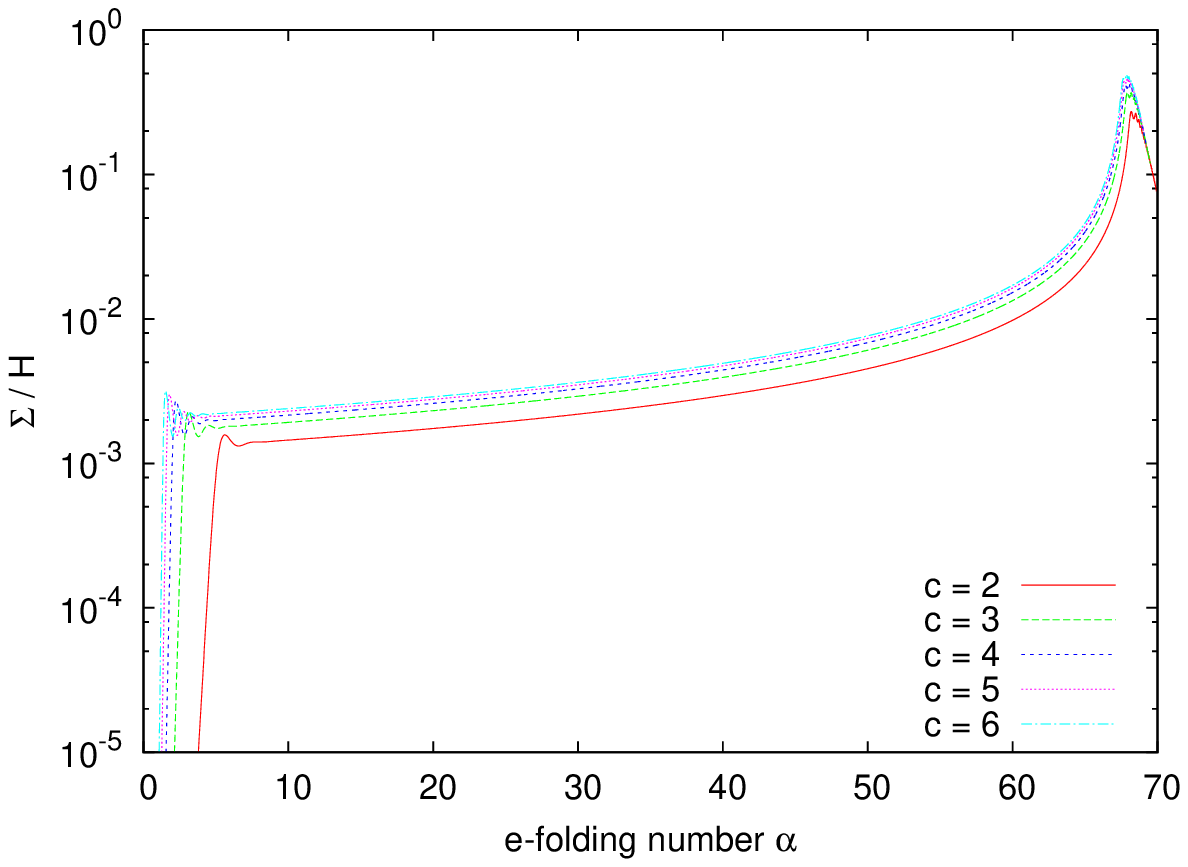}
\includegraphics[angle=0,width=75mm, height=65mm]{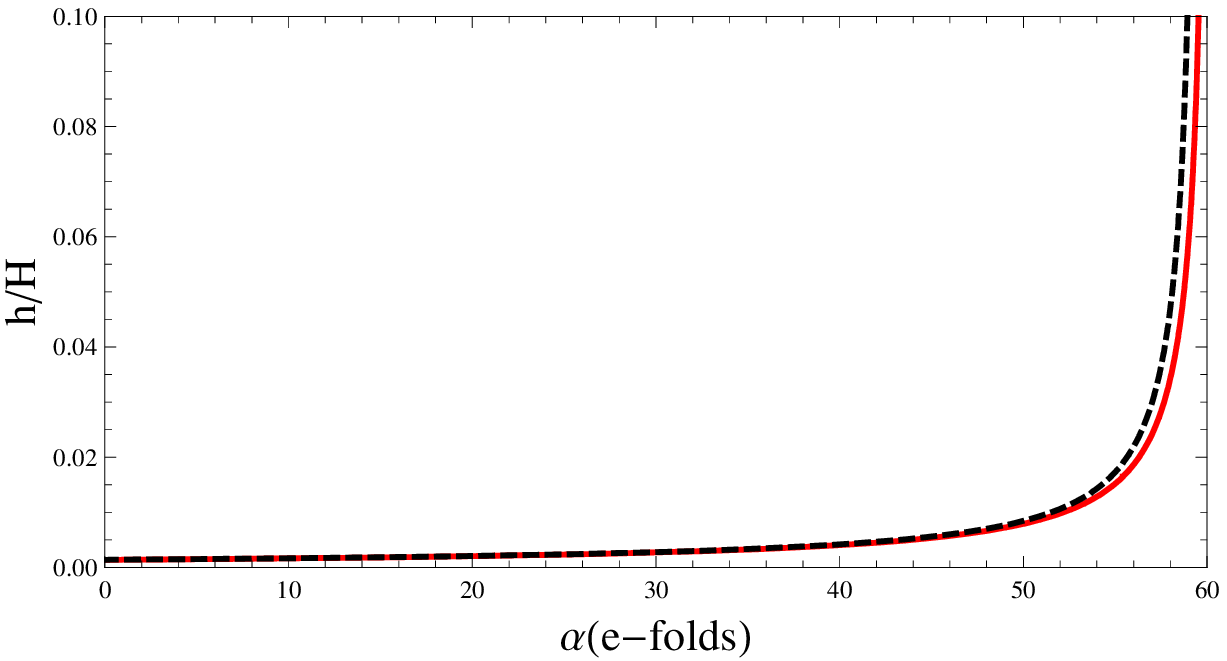}
\caption{In the left panel we have the evolutions of
Hubble-normalized shear $\frac{\Sigma}{H}:=\frac{\dot\sigma}{H}$ for
various $c$ values when $\sqrt{c}\varphi_i=17$. As we see, growing
during inflation and becoming of the order $\epsilon$,
$\frac{\Sigma}{H}$ saturates the upper-bound value. This figure is
taken from \cite{Kanno}. Then, in the right panel and for a
different initial values and parameters, we have the evolution of
Hubble-normalized shear $\frac{h}{H}:=\frac{\dot\sigma}{H}$ during
inflation when $c=2$ and $=\varphi_i=11M_{Pl}$. Again, anisotropies
grow during inflation and eventually saturate their upper bound
value. This figure is taken from \cite{Himmetoglu}.}\label{Jiro-fig}
\end{figure}

In \cite{Kanno}, the evolution of the Hubble-normalized shear ${\dot\sigma}/{H}$ for cases with $c>1$ has been calculated numerically.
They found two slow-roll phases and as expected, the
Hubble-normalized shear grows during inflation. In
Fig.\ref{Jiro-fig} left panel, we present their results for
${\dot\sigma}/{H}$ which is plotted for various values of parameter
$c$ under the initial condition $\sqrt{c}\varphi_i=17$. As wee see,
in the first slow-roll phase, the solutions show a rapid growth.
During the second phase it still grows but with a slower growth
rate. In the second slow-roll phase, in which $\ddot\sigma\ll H
\dot\sigma$, we approximately have %
\be
\frac{\dot\sigma(t)}{H}\simeq\frac{2}{3}\frac{\pi(t)}{H^2},
\ee
i.e.
the Hubble-normalized shear grows as a result of the growth of
Hubble-normalized anisotropic stress during the inflation. As has been discussed in \cite{Himmetoglu}, it is possible to choose the initial conditions and parameters such that the model has only one phase during about 60 e-folds of inflation. This, as shown in the right panel of Fig.\ref{Jiro-fig},
happens for  $\varphi_i=11M_{Pl}$ and $c=2$. As we see, the Hubble normalized anisotropy grows during
inflation and gradually saturates its upper bound value and becomes
of the order $\epsilon$.

 These
models exhibit an anisotropic attractor with non-zero, nevertheless
small $\dot\sigma/H$. As depicted in Fig.1, at the end of inflation
${\dot\sigma}/{H}\sim \epsilon$, saturating our bound
\eqref{upperSigma}. We comment that although the two-phase model of \cite{Kanno} does not strictly satisfy our slow-roll conditions for the two phases, since the second phase lasts long enough (\emph{cf.} Fig. 1), this model obeys our general arguments and the upper bound.

The above analysis has been extended for more general Bianchi models in \cite{Hervik} and for cases with more scalar fields and non-Abelian gauge fields \cite{Sodas, Soda-extenstions}. In all these cases the anisotropy dynamics is in accord with our general analysis of section 3, and can saturate the anisotropy bound \eqref{upperSigma} toward the end of inflation.

\subsubsection{Non-Abelian gauge field inflation, gauge-flation}

In \cite{gauge-flation}, we introduced a novel inflationary scenario, \emph{non-Abelian gauge field inflation} or \emph{Gauge-flation} for short. In this model inflation is driven by {non-Abelian gauge field} minimally coupled to Einstein gravity. It was shown that
non-Abelian gauge field theory can provide the setting for constructing an isotropic and
homogeneous inflationary background. This was achieved noting that gauge fields are defined up to gauge transformations and that any non-Abelian gauge group has an SU(2) subgroup. The global part of this SU(2) subgroup can be consistently identified with the rotation group. In particular, this means that among the gauge field components $A_\mu^a$ ($\mu$ being the space-time  and $a$ the gauge index) we have turned on a specific combination which behaves as an scalar under rotation. This scalar combination is coupled to all the other components of the gauge field through gauge interactions and hence one may excite these other components either classically or quantum mechanically, violating the isotropy and rotational symmetry of the background FLRW trajectory.

As discussed in details in \cite{gauge-flation}, one particularly convenient
choice of the gauge-flation action is
\begin{equation}\label{The-model}%
S=\int d^4x\sqrt{-{g}}\left[\frac{R}{2}-\frac{1}{4}F^a_{~\mu\nu}F_a^{~\mu\nu}+\frac{\kappa}{384}
(\epsilon^{\mu\nu\lambda\sigma}F^a_{~\mu\nu}F^a_{~\lambda\sigma})^2\right]\,,
\end{equation}
where $\epsilon^{\mu\nu\lambda\sigma}$ is the totally antisymmetric tensor and
the gauge field strength tensor $F^a_{~\mu\nu}$ is given by
\be
F^a_{~\mu\nu}=\partial_\mu A^a_{~\nu}-\partial_\nu A^a_{~\mu}-g\epsilon^a_{~bc}A^b_{~\mu}A^c_{~\nu},
\ee
here $\mu,\nu, ...$ labeled the space-time indices and run from 0 to 3, while $a,b, ...$ labeled the indices of the SU(2) gauge symmetry algebra and run from 1 to 3.
Our action consists of a Yang-Mills term (which due to scaling invariance can not lead to an inflating system) as well as a specific $F^4$ term, $Tr(F\wedge F)^2$, which its contribution to the energy-momentum tensor has the equation of state
$\rho_{\!_{F^4}}=-P_{\!_{F^4}}$, perfectly suited for driving an almost de Sitter expansion.\footnote{As discussed in \cite{gauge-flation} and in more
details in \cite{gauge-flation-vs-chromo-natural}, from particle
physics model building viewpoint, a $Tr(F\wedge F)^2$ type term can be argued for, considering
axions in a non-Abelian gauge theory and recalling the axion-gauge field interaction
term $\mathcal{L}_{axion}\sim \frac{\varphi}{\Lambda} Tr(F\wedge F)^2$. Then, integrating out the \emph{massive axion field} $\varphi$ leads to an action
of the form we have considered.}

Stability of the FLRW inflationary trajectory in gauge-flation against small fluctuations
around the isotropic FLRW background was already studied and established in \cite{gauge-flation}. Then, in \cite{gauge-flation cosmic no-hair} which will be briefly reviewed here,
starting from Bianchi type-I cosmology and through analytical and numerical studies, it was shown that the isotropic FLRW inflation is an attractor of the dynamics and the anisotropies are damped within a few e-folds, in accord with the cosmic no-hair conjecture. To do so, let us start with fixing the temporal gauge $A^a_{~0}=0$, with the ansatz
\be \label{ansatz1}
A=A^a_{~i}T^adx^i=e^{-2\sigma(t)}\frac{a(t)\psi(t)}{\lambda^2(t)} T^1dx+
e^{\sigma(t)}\lambda(t)a(t)\psi(t)(T^2dy+T^3dz)\,,
\ee
where $T^a$ are the $SU(2)$ gauge group generators,
$ [T_a,T_b]=i\epsilon^c_{~ab}T_c$. Note that $\lambda^2(t)=1,\ \sigma(t)=0$  corresponds to the isotropic background of \cite{gauge-flation}.

It turns out that the equations take a simpler form once written in terms of $\phi$
\bea\label{field}
\phi(t)=a(t)\psi(t)\,.
\eea
Substituting
the ansatz and the axi-symmetric Bianchi metric into $T_{\mu\nu}$ for the gauge fields, we obtain $\rho(t)$, $P(t)$ and $\pi(t)$
as
\be\label{rho-P-pi}
\begin{split}
\rho=\rho_\kappa+\rho_{_{YM}}, \quad P=&-\rho_\kappa+\frac{1}{3}\rho_{_{YM}},\\
\pi(t)=\frac23(1-\lambda^6)\big[\frac{1}{\lambda^4}(\frac{\dot\phi}{a}-2(\dot\sigma+\frac{\dot\lambda}{\lambda})
\frac{\phi}{a})^2-\frac{1}{\lambda^2}\frac{g^2\phi^4}{a^4}\big]
&-3\lambda^2(\frac{\dot\lambda}{\lambda}+\dot\sigma)\big[2\frac{\dot\phi}{a}-(\dot\sigma+\frac{\dot\lambda}{\lambda})
\frac{\phi}{a}\big]\frac{\phi}{a},
\end{split}\ee
where $\rho_\kappa$ and $\rho_{_{YM}}$  are respectively contributions of $F^4$ and Yang-Mills terms
\bea\label{k-rho}
\rho_\kappa&=&\frac{3}{2}\frac{\kappa g^2\phi^4}{a^4}\frac{\dot{\phi}^2}{a^2},\quad
\\
\label{ym-rho}
\rho_{_{YM}}&=&\frac{3}{2}\left[\frac{1}{3\lambda^4}\big(\frac{\dot\phi}{a}-2(\dot\sigma+\frac{\dot\lambda}{\lambda})\frac{\phi}{a}\big)^2
+\frac{2\lambda^2}{3}\big(\frac{\dot\phi}{a}+(\dot\sigma-\frac{\dot\lambda}{\lambda})\frac{\phi}{a}\big)^2
+\frac{(2+\lambda^6)}{3\lambda^2}\frac{g^2\phi^4}{a^4}\right]\,,
\eea
and
$$
\dot\sigma=-\frac{\big(\lambda^{-4}(\lambda^6-1)\phi^2\dot{\big)}}{2a^2\big(3+\lambda^{-4}(2+\lambda^6)
\frac{\phi^2}{a^2}\big)}.$$
Note that $\rho_\kappa$ is only a function of $\phi(t)$ and not $\lambda(t)$.
As we see, in the isotropic case $\lambda^2=1$ ($\dot{\sigma}=0$ and $\dot\lambda=0$), $\pi(t)$ vanishes.

Assuming that the system undergoes a quasi-de Sitter slow-roll inflation, the dynamics of $\lambda$ and $\sigma$ was examined and shown, both analytically and numerically, that this system has two attractor solutions $\lambda\equiv\big(\frac{\psi_2}{\psi_1}\big)^\frac13=\pm1$ which regardless of the initial values of $\lambda$, all the solutions converge to them within a few e-folds \cite{gauge-flation cosmic no-hair}.
These \emph{two attractor branches}, which are physically identical due to parity and charge conjugation invariance of our gauge-flation action \eqref{The-model}, correspond to the isotropic quasi-de Sitter solutions. Thus, gauge-flation is globally stable with respect to the initial anisotropies and respects cosmic no-hair conjecture.
\begin{figure}[h]
\includegraphics[angle=0,width=75mm, height=65mm]{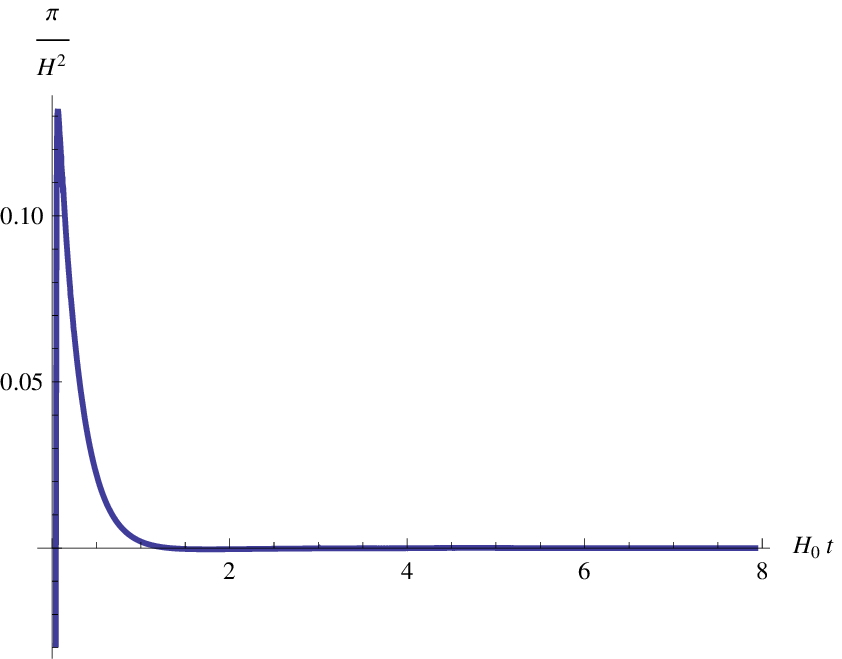}
\includegraphics[angle=0,width=75mm, height=65mm]{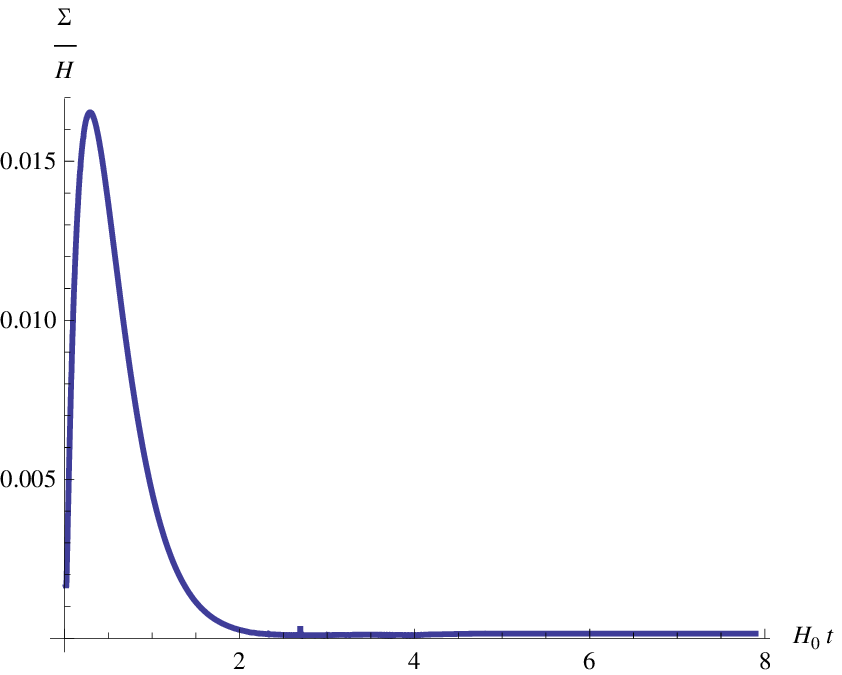}
\caption{Hubble-normalized anisotropic stress tensor, $\frac{\pi(t)}{H^2}$, and Hubble-normalized shear, $\frac{\Sigma(t)}{H^2}:=\frac{\dot\sigma(t)}{H^2}$ for a system with $\kappa=3.77\times10^{15}$,
$g=10^{-1}$, $\psi_0=0.6\times10^{-3}$, $\dot\psi_0=10^{-10}$, $\lambda_0=10$, $\dot\lambda_0=-3.6$. As we see
$\frac{\pi(t)}{H^2}$ increases (exponentially) for a very short time in the early stage of inflation (in small $H_0t$), saturating our upper bound \eqref{upperSigma}. Then, it is damped exponentially fast to its isotropic fixed point.}\label{gauge-flation}
\end{figure}

The slow-roll isotropic inflation is the attractor solution in our system, nonetheless, due to the vectorial nature in gauge-flation,
for very large and small initial values of $\lambda$ there is the possibility
for $\frac{\dot\sigma}{H}$ to saturate its upper bound value \eqref{upperSigma} for a very short lapse of time in early stages of inflation.
Both our numerical and analytical calculations reveals that in the extreme limits of $\lambda^6\gg1$ and $\lambda^6\ll1$, there is a region where anisotropy grows exponentially for a very short period, before getting exponentially damped to its isotropic fixed point. Although in these cases gauge-flation does not follow the strict dynamics indicated by the cosmic no-hair theorem \cite{Wald} for the very short period of time, the anisotropies are indeed damped within few Hubble times, in accord with the cosmic no-hair conjecture.

Although the analysis of \cite{gauge-flation cosmic no-hair} was carried out for Bianchi type-I model, a similar behavior for all Bianchi models is expected, once we assume having a quasi de Sitter slow-roll expansion phase. This is due to the fact that this inflationary phase is driven by $\rho_\kappa$  which dominates the $\rho_{\!_{YM}}$ and that the anisotropic stress $\pi(t)$ only receives contributions from the Yang-Mills term \eqref{rho-P-pi}, \eqref{k-rho} and  \eqref{ym-rho}.

\section{Concluding remarks}

In this work we extended Wald's cosmic no-hair theorem for general inflation setups, and in particular for slow-roll models. We proved an ``inflationary extended cosmic no-hair theorem'' which states that despite the fact that anisotropies can grow in time during inflation, their amplitude remains small and there is an upper bound on their amplitude.

 For Bianchi cosmological models and in the context of general relativity, we investigated the dynamics of anisotropies during inflation.
Considering a general imperfect fluid form for the energy-momentum tensor $T_{\mu\nu}$ we showed that the late time (after a few couple of e-folds) behavior of anisotropic tensor $\sigma_{\mu\nu}$ is governed by the anisotropic stress tensor $\pi_{\mu\nu}$;  if $\pi_i^{~j}(t)$ is vanishing or damped in time, as \eqref{diag} shows, the shear follows the same behavior as $\pi^i_{~j}$.
Choosing the orthonormal frame in which $T_{\mu\nu}$ is diagonal, $\sigma_{\mu\nu}$ may have off-diagonal elements in general. However, since these elements have no source term they are quickly damped, with time scale $H_0$.

Here we proved that inflation enforces an upper bound on
Hubble-normalized anisotropic stress ${\pi^i_{~j}(t)}/{H^2(t)}$,
proportional to the slow-roll parameter $\epsilon(t)$, which   can
in principle grow in time  during (slow-roll) inflation.  As a
result, inflationary dynamics allows the dimensionless anisotropies
${\sigma^i_i(t)}/{H(t)}$ to grow during the inflation, in contrast
to the cosmic no-hair conjuncture. Furthermore, assuming slow-roll
inflation, the dynamics of anisotropy does not necessarily follow
the slow-roll dynamics and they can generally  evolve quickly. This
effect was first noted in \cite{Kanno} in the context of a special
inflationary model with Bianchi type I metric in general relativity.
Although, inflation allows for the growth of the diagonal elements
of dimensionless anisotropy, as we showed here it puts an upper
bound $|\sigma^i_{~j}|/H\leq \frac83(\epsilon_0-\eta_0)$
\eqref{upperSigma} at the end of slow-roll inflation.

Our work was motivated in part by possible observational prospects and traces that a small but non-zero primordial anisotropy may have on the CMB, which may show up as statistical anisotropy in the CMB power spectrum. Currently, there are only bounds on the statistical anisotropy, for some studies in this direction see e.g. \cite{Sodas,Soda-extenstions,more-anisotropy}. These bounds will hopefully be improved by the upcoming cosmological observations and may be used as a tool to pin down on the model of inflation.
\vskip 5mm

\textbf{{Acknowledgement}}

We would like to thank Jiro Soda for his collaboration in the early stages of this work and for comments on the draft.
The work of AM is supported in part by the fund from Boniad-e Mell Nokhbegan of Iran.

\appendix

\section{Useful decomposition for energy momentum tensor of inflationary systems}\label{AsetUp}
Here, we prove that it is always possible to describe the energy momentum tensor $T_{\mu\nu}$ of any inflationary system (which satisfies DEC, but violates SEC) as
\be\label{decom}
T_{\mu\nu}=-\Lambda(t)g_{\mu\nu}+\mathcal{T}_{\mu\nu}\,,
\ee
where $\Lambda(t)\geq 0$, and $\mathcal{T}_{\mu\nu}$ which satisfies both SEC and WEC.

Assuming inflationary dynamics in a system, we have
$$\dot H+H^2\geq0\,,$$
which using \eqref{ray} gives the following inequality for the total energy-momentum tensor $T_{\mu\nu}$
\be\label{inflation-condition}
(T_{\mu\nu}-\frac12 g_{\mu\nu}T)n^\mu n^\nu+\sigma_{\mu\nu}\sigma^{\mu\nu}=-3(\dot H+H^2)\leq0.
\ee
As a necessary condition for the above inequality to be satisfied, $T_{\mu\nu}$ should violate SEC.\footnote{We comment that having inflation in generic anisotropic models demands a condition stronger than violation of SEC, it requires $(T_{\mu\nu}-\frac12 g_{\mu\nu}T)n^\mu n^\nu\leq -\sigma_{\mu\nu}\sigma^{\mu\nu}$, as indicated in \eqref{inflation-condition}.}
However, decomposing $T_{\mu\nu}$ as \eqref{decom}, one can choose $\Lambda(t)$ in such a way that
\be\label{L1}
\Lambda(t)\geq3(\dot H+H^2)+\sigma_{\mu\nu}\sigma^{\mu\nu},
\ee
and obtain
\be
(\mathcal{T}_{\mu\nu}-\frac12 g_{\mu\nu}\mathcal{T})n^\mu n^\nu\geq0.
\ee
That is, choosing $\Lambda(t)$ as \eqref{L1} one can obtain a $\mathcal{T}_{\mu\nu}$ which respects SEC.

On the other hand, for $\mathcal{T}_{\mu\nu}$ to satisfy WEC, we need $\mathcal{T}_{\mu\nu}n^\mu n^\nu\geq0$, which from \eqref{initialvalue} requires
\be\label{L2}
\Lambda(t)\leq3H^2+\frac12{}^{(3)}R-\frac12\sigma_{\mu\nu}\sigma^{\mu\nu}.
\ee
Then, demanding $\mathcal{T}_{\mu\nu}$ to satisfy both SEC and WEC, \eqref{L1} and \eqref{L2} imply that  $\Lambda(t)$ should be in the following
region
\be\label{L}
3(\dot H+H^2)+\sigma_{\mu\nu}\sigma^{\mu\nu}\leq\Lambda(t)\leq3H^2+\frac12{}^{(3)}R-\frac12\sigma_{\mu\nu}\sigma^{\mu\nu}.
\ee
In order to prove the validity of such a choice, one should show that the upper bound value for $\Lambda(t)$ is less than its lower bound value.
In the following by showing that the difference of the upper and lower bound values is a positive quantity, we prove the validity of \eqref{L}.

The total energy-momentum tensor $T_{\mu\nu}$ satisfies DEC which implies that $T_{\mu\nu}n^\mu n^\nu+\frac13T_{\mu\nu}h^{\mu\nu}\geq0$. We hence  have
$$\mathcal{T}_{\mu\nu}n^\mu n^\nu+\frac13\mathcal{T}_{\mu\nu}h^{\mu\nu}\geq0\,,$$
which recalling \eqref{initialvalue} and \eqref{ray} yields
\be\label{quantity}
-3\dot H-\frac32\sigma^{\mu\nu}\sigma_{\mu\nu}+\frac12{}^{(3)}R\geq0\,.
\ee
The above is exactly the difference of the upper and lower values in \eqref{L}. We have thus proved the validity of the statement made in the beginning of this appendix.

We note that as is implicit in the above, SEC and WEC conditions does not uniquely specify $\Lambda(t)$. One may hence assume extra conditions on $\mathcal{T}_{\mu\nu}$. One such choice is to take $\mathcal{T}_{\mu\nu}$ to be stiff matter, that is a matter with $\rho=P$ (with possibly non-zero anisotropic stress $\pi_{\mu\nu}$). This freedom and various choices of $\Lambda(t)$ does not lead to a stronger bound on the anisotropy.

\section{{Wald's theorem assumptions and inflationary models}}
\label{AWald}

Consider inflationary models in which we have exponential expansion in a finite period of time and inflation eventually ends.
These systems do not respect conditions of Wald's theorem \cite{Wald}.  Systems which are described by Wald's theorem evolve toward the de Sitter solution and inflation will never stops in them. 
Here we review in more technical details exactly how models of inflation fail fulfilling  assumptions of Wald's cosmic no-hair theorem.

To this end, we employ Wald's notations  \cite{Wald} where total energy momentum $T_{\mu\nu}$  \eqref{einst} was decomposed in terms of a positive cosmological \emph{constant} $\Lambda_0$, and an extra term $\tilde T_{\mu\nu}$
\be
T_{\mu\nu}=-\Lambda_0g_{\mu\nu}+\tilde T_{\mu\nu}\,,
\ee
with $\tilde T$ satisfying DEC and SEC. This latter is not respected in inflationary models. In terms of our notations, \be\label{B2}
\tilde T_{\mu\nu}=-(\Lambda(t)-\Lambda_0)g_{\mu\nu}+\mathcal{T}_{\mu\nu}\,.
\ee
As we find from the above equation for any given $T_{\mu\nu}$,  there is an ambiguity in defining $\tilde T_{\mu\nu}$, unless the value of $\Lambda_0$ is also specified. In order to resolve this ambiguity, we choose $\Lambda_0$ to be
\be\label{initialdecom1}
\Lambda_0=3H^2|_{_{\Sigma_0}}\,,
\ee
 which using \eqref{initialvalue},  determines initial value of $n^\mu n^\nu\tilde T_{\mu\nu}$
\be\label{initialT}
n^\mu n^\nu\tilde T_{\mu\nu}|_{_{\Sigma_0}}=\frac12{}^{^{(3)\!\!}}R|_{_{\Sigma_0}}-\frac12\sigma_{\mu\nu}\sigma^{\mu\nu}|_{_{\Sigma_0}},
\ee
where $\Sigma_0$ represents spatial hypersurface of initial time at the beginning of inflation.
Recalling  that for all Bianchi models expect type IX, ${}^{^{(3)\!\!}}R\leq0$, equation \eqref{initialT} implies that $\tilde T_{\mu\nu}|_{_{\Sigma_0}}$ does not respect the dominant energy condition in these models.

Noting that $H^2$ is a decreasing quantity in inflationary systems, and using \eqref{initialvalue}, \eqref{B2} and \eqref{initialT}, one finds out that
\begin{itemize}
\item{for all Bianchi cosmological models expect type IX, we have $n^\mu n^\nu\tilde T_{\mu\nu}\leq0$ during the inflation, which implies that $\tilde T_{\mu\nu}$ violates DEC.}
\item{In addition, in case of having inflation in Bianchi type IX in which ${}^{^{(3)\!\!}}R\geq0$, the spatial curvature is rapidly damped and after a few e-folds, we have $n^\mu n^\nu\tilde T_{\mu\nu}\leq0$, which again implies $\tilde T_{\mu\nu}$ violates DEC.  }
\item Since during inflation $H^2$ is a decreasing quantity, one can show that it is not possible to find a $\Lambda_0>0$ in such a way that the resulting $\tilde T_{\mu\nu}$ satisfies SEC and DEC for the whole period of inflation.
\end{itemize}

Note that although $\tilde T_{\mu\nu}$ eventually violates DEC in inflationary systems, the total energy-momentum tensor $T_{\mu\nu}$, always satisfies DEC and the energy density of the system, $T_{\mu\nu}n^\mu n^\nu$, is a positive definite quantity.
To summarize, we showed that during inflation, if $\tilde T_{\mu\nu}$ satisfies SEC it will necessarily violate DEC, which is against the requirements of Wald's cosmic no-hair theorem.


\end{document}